# Design and Simulation of Symmetric Nanostructures Using Two-beam Modulated Interference Lithography Technique

A. Alfred Kiruba Raj[2], D. Jackuline Moni[2], D. Devaprakasam[1*]

[1]Department of Nanosciences and Technology, School of Nanosciences and Technology

[2]Department of Electronics and Communication Engineering
Karunya University, Coimbatore 641114, Tamil Nadu, INDIA

*Abstract*— Laser Interference Lithography (LIL) is a versatile fabrication method for patterning sub-micron structures in arrays covering large areas. It is a facile and fast mask-less lithography process to produce large area periodic patterns. The objective in this study is to simulate and design periodic and quasi-periodic 1D, 2D and 3D nanostructures using two-beam interference technique. We designed and simulated periodic and quasi-periodic structures by two-beam interference patterning using a MATLAB program by varying angle of incidence, wavelength and geometry. The generated patterns/features depend on the interference intensity, wavelength, slit separation and angle of incidence. Using this technique, we can achieve potentially high-volume of uniformity, throughput, process control, and repeatability. By varying different input parameters, we have optimized simulated patterns with controlled periodicity, density and aspect ratio also it can be programmed to obtain images of interference results showing interference intensity distributions.

## I. Introduction

Laser Interference Lithography (LIL) produces pattern structures when two or more coherent light beams intersect. The patterned array features have series of fringes with minima and maxima. LIL produces periodic large array structure with high resolution over extremely large field sizes about ~1 m [1]. This technique has been successfully applied to manufacture periodic structures of sensors and devices.

The advantage on LIL is simple, more effective to fabricate 1D, 2D & 3D structure. The coherent beams are controlled by computer interface for varying the wavelength λ. LIL can be used as mask-less lithography [2-4]. The current method of Photo-resist uses single exposures [6] which are used for high resolution where the simplest interference patterns are realized. The area coverage is about 0.25 cm$^2$ also the exposed area is limited by mirror size [5]. The comparison study on different lithography techniques is shown in table I. From the table, the important parameters wavelength, exposing particle, types of optics used, mask type, resolution limitation and depth of focus are tabulated.

TABLE I. COMPARISONS OF DIFFERENT TYPES OF LITHOGRAPHY'S

| Parameter | U.V | X-Ray | E-beam | I-beam | *Laser interference |
|---|---|---|---|---|---|
| Effective wavelength | 11 nm | 0.1 nm | 12 nm | 0.1 nm | 1064nm |
| Exposing particle | Photons | Photons | Electrons | Ions (Protons) | Photons |
| Type of optic | Reflective | Reflective | Electromagnetic | Electromagnetic | None |
| Mask type | Reflective | Transmission | None | Transmission | None |
| Resolution limit | 45 nm | 30 nm | 22 nm | 2 nm | 64nm |
| DOF | 1100 nm | Large | Large | Large | Adjustable |

In this work, Section II deals about configuration of optical interference lithography systems, Section III deals about Proposed Work – DBI Technique in MATLAB, Section IV deals about simulation of DBI Technique, Section V deals about experimental result and its discussion and Section VI on conclusion.

## II. CONFIGURATION OF OPTICAL INTERFERENCE LITHOGRAPHY SYSTEMS

### A. Two-beam exposure

The principle behind the two beam exposure technique is the coherent light is split into two or more beams; they recombine to produce interference pattern structures. The resultant intensity is observed in the patterns of dark and bright fringes.

$$I_r = I_1 + I_2 + 2\sqrt{I_1 \times I_2} \times \cos\phi$$

## B. Construction of DBI technique:

Laser acts as light source (it may be Argon laser, HeCd, YAG laser, Nd: YAG laser, etc...[6, 7]), the beams are guided by the mirrors and is spitted into two half's by a beam splitter as shown in Figure 1.1 (a, b). The beam from beam splitter is passed through attenuator and polarizer to ensure that both beams have same power with specific polarization. The beams are expanded and filtered by spatial filter which removes high frequency noise. The filtered beams intersect and produces interference pattern.

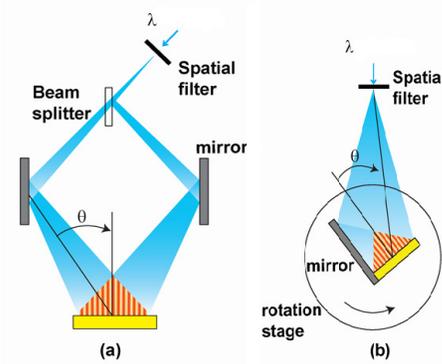

Figure 1.1: (a, b) Construction of Two beam exposure [9]

## III. Proposed Work – DBI Technique

The proposed DBI technique can produce more complex micro/nano pattern structures than conventional method. The expected area coverage in DBI is more compared with single exposure method. Lloyds Mirror Interferometer (LMI) also produces patterns but there requires the phase correction. The main features of the proposed method over conventional lithography is 1) Mask less lithography, 2) Depth of focus is more, 3) More number of complex micro/nano patterns can be obtained, 4) Direct writing is possible. To obtain good line width control, the latent image remains in focus through the depth of the resist layer. The depth of focus (DOF) or δ of an optical system is given by

$$DOF = \frac{K_2 \lambda}{NA}$$

Where $\lambda$ is the wavelength of the laser beam, NA is the numerical aperture of optic used and $K_2$ is the process dependent constant.

### A. Mathematical modelling of DBI technique

Let us assume that the source is monochromatic of wavelength $\lambda$. Two narrow pinholes $S_1$ and $S_2$ are kept at equal distance from s i.e., $sS_1 = sS_2$ as shown in Figure 2.1. Consider a point 'O' on the screen, which is equidistant from $S_1$ and $S_2$. The maximum intensity occurs at point 'O'.

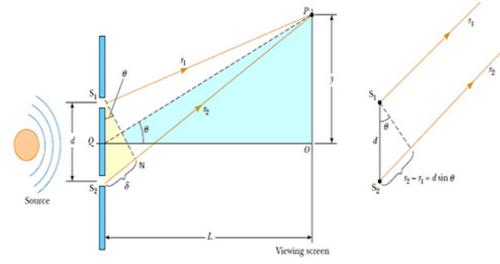

Figure 2.1: Modelling of DBI Technique [9]

Consider 'p' at a distance 'y' from 'O'. The light intensity on the screen at point P is the resultant of the light from both the slits. A wave from lower $S_1$ travels farther than a wave from the upper slit by the amount of 'dsinθ'. This distance is called path difference 'δ'.

$$\delta = r_1 - r_2 = d\sin\theta \qquad (1)$$

$$\delta = S_2 N \qquad (2)$$

From $\Delta S_1 S_2 N$,

$$\Rightarrow \frac{S_2 N}{S_1 S_2} = \sin\theta,$$
$$\Rightarrow S_2 N = d\sin\theta,$$
$$\Rightarrow \delta = d\sin\theta$$

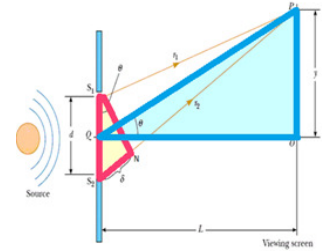

From $\Delta QOP$,

When $\theta$ is small, $\theta = \frac{y}{L}$

Hence $\sin\theta = \tan\theta = \theta = \frac{y}{L}$

Therefore Path difference, $\delta = d\sin\theta = d\theta = d\frac{y}{L}$

$$\Rightarrow \delta = d\frac{y}{L} \quad \text{and} \quad phase\,difference = \frac{2\pi}{\lambda} \times \delta$$

For constructive interference, the path difference is 0 or some integral multiple of the wavelength.

$$\delta = d\sin\theta = m\lambda \quad \text{Where} \quad m = 0, \pm 1, \pm 2, \ldots \qquad (3)$$

For destructive interference,

$$\delta = d\sin\theta = \left(m + \frac{1}{2}\right)\lambda \quad \text{Where} \quad m = 0, \pm 1, \pm 2, \ldots \qquad (4)$$

Similarly from $\Delta S_1S_2L$,

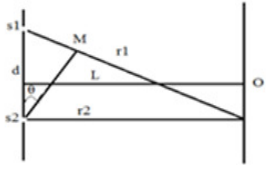

$$\sin\theta = \frac{S_1L}{d} \quad (5)$$

$$d\sin\theta = S_1L = \delta \quad (6)$$

By combining the equations (1) and (6), we get

$$\delta = d\sin\theta + d\sin\theta = 2d\sin\theta \quad (7)$$

Where $\delta$ is the numerical aperture.

### B. Depth Of Focus (DOF)

Depth of focus is defined as the range of focus errors that a process can tolerate and still give acceptable lithographic results. $DOF = \frac{K_2\lambda}{(\delta)^2} = \frac{K_2\lambda}{(d\sin\theta)^2}$

$$\Rightarrow DOF = \frac{K_2\lambda}{d^2\sin^2\theta} \quad (8)$$

Where $\lambda$ is the wavelength of the laser light, d is the distance between two slits, $K_2$ is a process dependent constant.

### C. Intensity Derivation For Two Beam

Consider the two beams of light waves Y1 and Y2,

where

Y1 = $E_o \sin(kx-wt)$ (9)

Y2 = $E_o \sin(kx-wt+\varphi)$ (10)

In order to get the resultant component, we have to add both the Y components (Vector Addition). Therefore the amplitude of these waves is identical but there will be a difference in the phase,

$$\phi = \frac{2\Pi d\sin\theta}{\lambda} \quad (11)$$

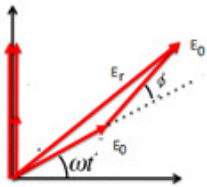

Resultant electric field is $E_r$, the intensity is simply proportional to the square of the amplitude which is shown as phasor diagram. Resultant wave with amplitude $E_r$ and the phase shift relative to the original wave is $\beta$.

$$\therefore \beta = \frac{1}{2}\varphi$$

$$E_r = 2(E_0\cos\beta) \quad (12)$$

$$= 2E_0\cos\frac{\varphi}{2}$$

$$E_r = 2E_0\cos\left(\frac{\pi d\sin\theta}{\lambda}\right) \quad (13)$$

Intensity is the square of the amplitude, so

$$I = \left[2E_0\cos\left(\frac{\pi d\sin\theta}{\lambda}\right)\right]^2$$

$$= 4I_0\cos^2\left(\frac{\pi d\sin\theta}{\lambda}\right)$$

$$\frac{I}{I_0} = 4\cos^2\left(\frac{\pi d\sin\theta}{\lambda}\right)$$

Thus the resultant intensity of two beam is given by

$$\frac{I}{I_0} = 4\cos^2\left(\frac{\pi d\sin\theta}{\lambda}\right) \quad (14)$$

## IV. Simulation of DBI Technique in MATLAB

Arrays of dots or holes on the surface of material are produced using dual beam interference. To illustrate the interference pattern generated by two beam interfering coherent beams, a theoretical simulation using MATLAB software was performed. The Figure 3.1 (a) & (b) shows distribution of laser intensity corresponding to the interference of two laser beams ($\lambda$=700nm, 355nm). We found that, as the wavelength (nm) gets varied with respective with the slit width (mm), the distribution of laser intensity and the interference pattern also changes.

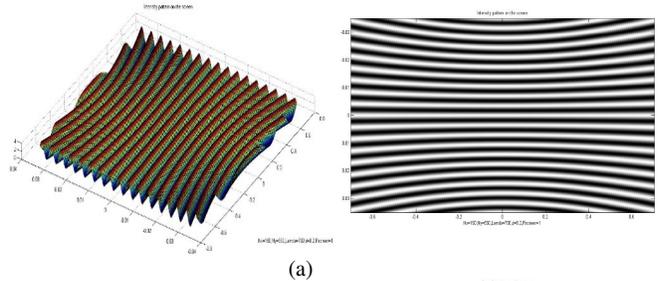

(a)

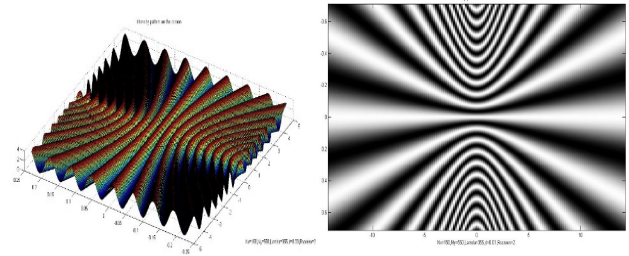

(b)

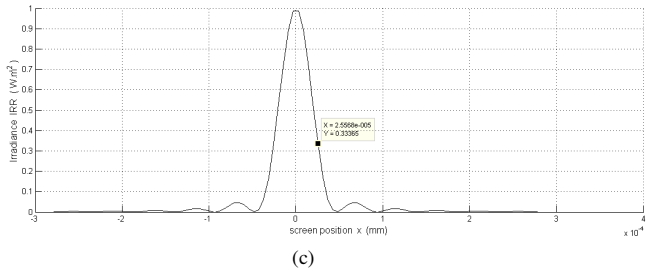

(c)

Figure 3.1: (a), (b) are the intensity pattern over the screen and (c) shows the wave pattern formation.

*A. Analyzing the patterns by varying the distance between the slits*

- Two Beam at d=0.25mm:

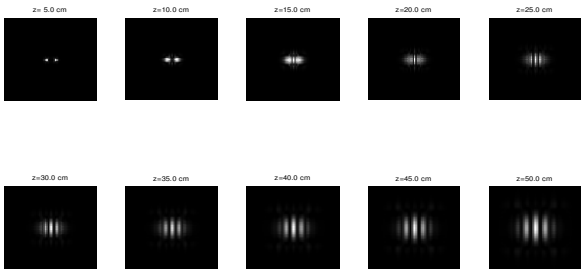

- Two Beam at d=0.5mm:

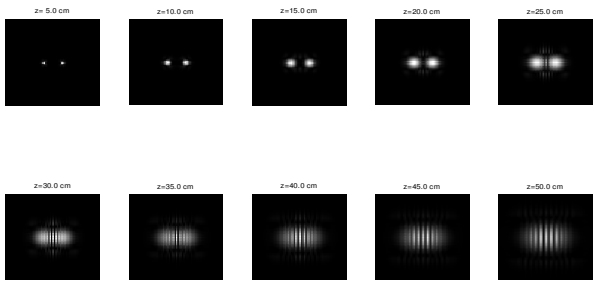

- Two Beam at d=1mm:

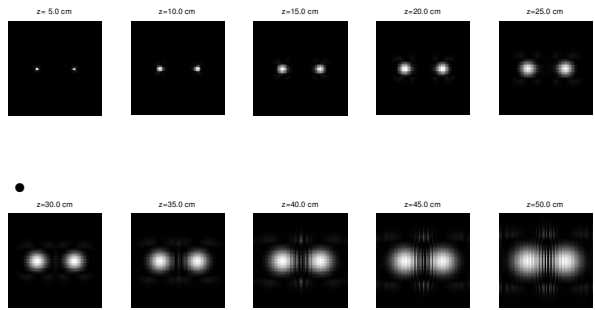

- Two Beam at d=2mm:

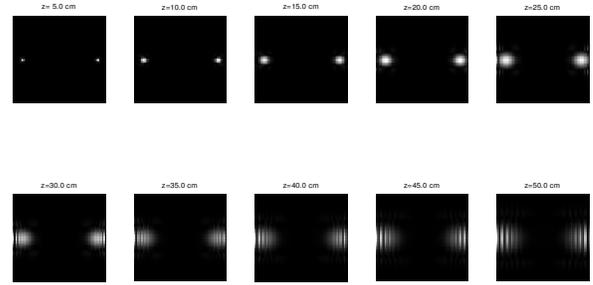

Figure 4.1: Shows the variation in slit separation (mm) with respective to the beam intersects.

From the above Figure 4.1, we conclude that, as the distance between the slits increases, the interference will be more as they are directly proportional.

## V. Experimental Result and its discussion

In Dual Beam Interferometer (DBI), we varied slit width, incidence angle, slit separation and depth of penetration. We observed, as slit width increases from mina to maxima, the beam incidence at screen also increases. On varying slit separation with depth of penetration, it was observed that lesser the slit separation results in higher penetration of beam in the substrate as shown in table II and III.

TABLE II.   SHOWS THE SLIT WIDTH (MM) VERSUS INCIDENCE ANGLE

| slit width (mm) | Incidence angle (degree) |
|---|---|
| 0.6 | 0.045 |
| 0.9 | 0.068 |
| 1.3 | 0.099 |
| 5.9 | 0.45 |
| 10 | 0.76 |
| 50 | 3.83 |
| 100 | 8.1 |
| 500 | 35.44 |
| 800 | 48 |
| 1000 | 63.43 |
| 5000 | 45 |

TABLE III.   SHOWS THE SLIT SEPARATION VERSUS DEPTH OF PENETRATION OF BEAM ANGLE

| | Depth of penetration of beam | | |
|---|---|---|---|
| Slit Separation(mm) | Top value | Middle value | Bottom value |
| 0.02 | 1.48 | 2.74 | 4.44 |
| 0.1 | 1.06 | 1.9 | 3.6 |
| 0.2 | 0.62 | 1.48 | 2.32 |
| 0.3 | 0.62 | 1.06 | 1.48 |
| 0.5 | 0.2 | 0.62 | 1.06 |
| 0.6 | 0.2 | 0.63 | 1.06 |

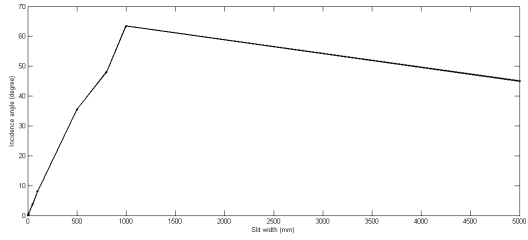

Figure 4.1: Graphical representation over the slit width (mm) versus Incidence angle (degree)

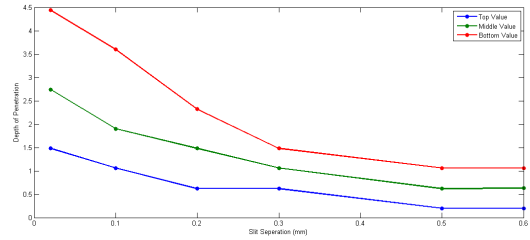

Figure 4.2: Graphical representation over the slit separation (mm) depth of penetration

### B. Calculation of depth of focus:

TABLE IV.  SHOWS THE CALCULATION OF DEPTH OF FOCUS

| Wavelength (nm) | $DOF = K_2 \dfrac{\lambda}{(NA)^2}$ |
|---|---|
| 355 | 78.6 nm |
| 595 | 131 nm |
| 779 | 172 nm |
| 1064 | 235 nm |

From Table IV, $K_2$ is the process parameter is around 0.2. The numerical aperture (NA) is around 0.95. The depth of focus (DOF) of focus is calculated for varying wavelengths. For 355nm, the depth of focus is

$$DOF = K_2 \dfrac{\lambda}{(NA)^2} \quad (15)$$

$$= \dfrac{0.2 \times 355 \times 10^{-9}}{(0.95)^2} = 78.6 nm$$

## VI. CONCLUSION

The formation of symmetric micro/nano pattern structures are developed using dual beam interference method. We have simulated the 2D and 3D pattern structure using MATLAB. In this study, we optimized the 2D and 3D pattern structure by varying the angle of incidence, wavelength, slit width and slit separation. We obtained several fridge patterns in accordance with the varying slit width and slit separation as shown in Figure 3.1 (a &b). By analyzing the simulated structure, we observed that the maxima of the interference patterns mostly depend on the angles of incidence at 45°. On increasing the number of beams, more complicated patterns can be formed over the substrate. Figure 4.2 shows variation in slit separation with Depth of penetration, in which it is observed as slit width increases from mina to maxima correspondingly the angle at which the beam incidence at screen also increases.

### Acknowledgement

We thank DST-Nanomission, Government of India and Karunya University for providing the financial support to carry out the research. We also thank the department of Nanosciences and technology and the department of electronics and communication engineering for the help and support to this research.

### References

[1] Henk van Wolferen and Leon Abelmann,"Laser interference lithography," in Lithography: Principles, Processes and Materials.2011 Nova Science Publishers, Inc., pp.133-148.

[2] S.R.J.Brueck,"Optical and Interferometric Lithography-Nanotechnology Enablers," proceedings of the IEEE, VOL.3, NO.10, October 2005.

[3] S.R.I. Gabran, R.R. Mansour, M.M.A.Salama, "Maskless Pattern Transfer Using 355 nm Laser, "in Optics and Lasers in Engineering", 50(2012) 710-716

[4] Yung-Lang Yang, Chin-Chi Hsu. Study on wetting properties of periodical nanopatterns by a combinative technique of photolithography and laser interference lithography. Applied Surface Science (2010):3683-3687

[5] Lei Zhang, Tielin Shi, Carbon nanotube integrated 3-dimensional carbon microelectrode array by modified SU-8 photoresist photolithography and pyrolysis. Thin Solid Films(2011):1041-1047

[6] S.R.I. Gabran, R.R. Mansour, Maskless pattern transfer using 355 nm laser. Optics and Lasers Engineering (2012):710-716

[7] Ren Yang, Steven A.Soper, Wanjun Wang. "A new UV lithography photoresist based on composite of EPON resins 165 and 154 for fabrication of high-aspect-ratio microstructures". Sensors and Actuators A 135(2007):625-636.

[8] By Chiara Ingrosso, Vahid Fakhfouri, Marinella Striccoli, Angela Agostiano, Anja Voigt, Gabi Gruetzner, M. Lucia Curri, and Juergen Brugger. An Epoxy Photoresist Modified by Luminescent Nanocrystals for the Fabrication of 3D High-Aspect-Ratio Microstructures. Adv. Funct. Mater. (2007): 2009–2017.

[9] Y.Zabila, M.Perzanowski, A. Dobrowolska,"Direct Laser Interference Patterning: Theory and Application" University of Science and Technology, al. Mickiewicza 30, 30-054 Kraków, Poland,

[10] S. Cabrini A. Carpentiero R. Kumar "Focused ion beam lithography for two dimensional array structures for photonic applications",

[11] Keiper, Kuntze, Drilling of Glass by Excimer Laser Mask projection Technique. Laser Application (2000):189-193

[12] Ingemar Eriksson, "the monitoring of a laser beam", department of information technology and media, mid Sweden university

[13] Chris A mack, "Depth of Focus" in the lithography tutor FINIL Technologies, Austin, Texas

[14] S. M. Schultz," Using MATLAB to help teach Fourier optics", Dept. of Electrical and Computer Engineering, Brigham Young University, Provo, UT USA 84602

[15] D. Akoury"The Simplest Double Slit: Interference and Entanglement in Double Photoionization of H2".